\documentclass[twocolumn,twoside]{revtex4}
\usepackage{graphicx}
\usepackage{fancyhdr}
\usepackage{amssymb}
\usepackage{amsmath}
\usepackage{amsfonts}
\newcommand{\tautau}{{\tau^+\tau^-}}
\newcommand{\gamgam}{{\gamma\gamma}}
\newcommand{\bbbar}{{b \bar b}}
\newcommand{\mev}{\,\mathrm{MeV}}
\newcommand{\gev}{\,\mathrm{GeV}}
\newcommand{\tev}{\,\mathrm{TeV}}
\newcommand{\tb}{{\tan\beta}}

\newcommand\SW{s_\mathrm{w}}
\newcommand\CW{c_\mathrm{w}}
\newcommand\MW{M_W}
\newcommand\MWCDF{\ensuremath{M_W^{\rm CDF}}}
\newcommand\MWexp{\ensuremath{M_W^{\rm exp}}}
\newcommand\MWexpnew{\ensuremath{M_W^{\rm exp,new}}}
\newcommand\MZ{M_Z}
\newcommand\MWSM{\MW^{\rm SM}}

\newcommand\MWMSSM{\MW^{\rm MSSM}}

\newcommand\sweff{\sin^2 \theta_\mathrm{eff}}

\newcommand{\seff}[1]{\sin^2\theta_{\rm eff}^{#1}}
\newcommand{\MH}{M_H}

\def\amumssm{\ensuremath{a_\mu^{\rm MSSM}}}
\def\gmin2{\ensuremath{(g-2)_\mu}}
\def\amu{\ensuremath{a_\mu}}
\newcommand\neu[1]{\tilde{\chi}^0_{#1}}
\newcommand\cha[1]{\tilde{\chi}^\pm_{#1}}
\newcommand{\Slpm}{\tilde l}

\newcommand\al{\alpha}
\newcommand\be{\beta}
\newcommand\ga{\gamma}
\newcommand\De{\Delta}
\newcommand\si{\sigma}

\newcommand{\mt}{m_t}

\newcommand\refeq[1]{Eq.~(\ref{#1})}

\newcommand\refse[1]{Sect.~\ref{#1}}

\newcommand\citere[1]{Ref.~\cite{#1}}
\newcommand\citeres[1]{Refs.~\cite{#1}}
\def\reffi#1{\mbox{Fig.~\ref{#1}}}

\newcommand\LB{\left[}
\newcommand\RB{\right]}
\newcommand\LV{\left\{}
\newcommand\RV{\right\}}

\begin{document}

\title{CDF Measurement of \boldmath{$\MW$}: Theory implications}

%

\author{S.~Heinemeyer}
\affiliation{Instituto de F\'isica Te\'orica UAM-CSIC, Cantoblanco, 28049,
      Madrid, Spain}

\begin{abstract}
The CDF collaboration recently reported a measurement of the $W$-bosos
mass, $M_W$, showing a large positive
deviation from the Standard Model
(SM) prediction.
The question arises whether extensions of the SM exist that can
accommodate such large values, and what further phenomenological
consequences arise from this.
We give a brief review of the implications of the new CDF measurement
on the SM, as well as on Higgs-sector extensions. In particular, we
review the compatibility of the $\MW$ measurement of CDF with excesses
observed in the light Higgs-boson searches at $\sim 95 \gev$, as well
as with the Minimal Supersymmetric Standard Model in conjunction with
the anomalous magnetic moment of the muon, \gmin2. 
\end{abstract}

\maketitle

\thispagestyle{fancy}


\section{Introduction}
\label{sec:intro}

The mass of the $W$~boson can be predicted from muon decay, which relates
$\MW$ to three extremely precisely measured quantities: the Fermi
constant, $G_\mu$, the fine structure constant, $\al$, and the mass of the
$Z$~boson, $\MZ$. Within the SM and many extensions of it
this relation can be used to predict $\MW$ via the expression
\begin{align}
\label{eq:mwpred}
  & \MW^2 = \MZ^2 \times \\
  &\LV \frac{1}{2} +
\sqrt{\frac{1}{4} - \frac{\pi\,\al}{\sqrt{2}\,G_\mu\,\MZ^2}
\LB 1 + \De r(\MW, \MZ, \mt, \ldots) \RB } \RV, \nonumber
\end{align}
where the quantity $\De r$ is zero at lowest order.
It comprises loop corrections to muon decay in the considered model, where the
ellipsis in \refeq{eq:mwpred} denotes the specific particle content of the
model.

The SM prediction for $\De r$ includes contributions at the complete
one-loop~\cite{Sirlin:1980nh,Marciano:1980pb} and the complete
two-loop level,
as well as partial higher-order corrections up to four-loop
order (see, e.g., \citere{PDG2022} for a review).
This yields a prediction of
\begin{align}
\MWSM = 80.357 \pm 0.004 \gev~,
\label{mwsm}
\end{align}
where the uncertainty originates from unknown higher-order corrections.
This value is in agreement at the $\sim 2\,\si$ level with the current
PDG average~\cite{PDG2022}
\begin{align}
\MW^{\rm PDG} &= 80.377 \pm 0.012 \gev\,.
\label{exp-cwa}
\end{align}

Recently the CDF collaboration reported a new measurement using their
full data set of $8.8$\,fb$^{-1}$ of~\cite{CDF:2022hxs}, 
\begin{align}
\MWCDF &= 80.4335 \pm 0.0094 \gev\,,
\label{cdf-new}
\end{align}
which deviates from the SM prediction by $7.0\,\si$. Combining this
new value with other measurements of the Tevatron and LEP (but not the LHC)
yields,
\begin{align}
\MW^{\rm Tev+LEP} &= 80.4242 \pm 0.0087 \gev\,.
\label{tev-lep}
\end{align}
A possible new world average
including the recent CDF measurement~\cite{CDF:2022hxs}, is roughtly
given by~\cite{PDG2022}%
\footnote{It should be noted that the values given so far in
\citere{PDG2022} are rather approximate. }%
, 
\begin{align}
  \MWexpnew &\approx  80.417 \pm 0.018 \gev~.
\label{exp-nwe}
\end{align}
The enlarged uncertainty of \MWexpnew\ reflects the fact that the new CDF
measurement is not well compatible with previous experimental
results. 
In the future it will be mandatory to assess the compatibility
of the different measurements of $\MW$ and to carefully analyze 
possible sources of systematic effects. However, naturally the
question arises whether extensions of the SM exist that can
accomodate such large values, and what further phenomenological
consequences arise from this. 

Here we will briefly review why the SM is incompatible with $\MWCDF$,
and how extensions of the SM Higgs sector can accomodate values
substantially above the SM prediction. We furthermore review whether
the electroweak (EW) sector of the Minimal Supersymmetric Standard
Model (MSSM), while 
being in agreement with \gmin2\ can (or cannot) give rise to such high
values of $\MW$, and finally comment on contributions from scalar
tops/bottoms.


\section{Incompatibility of the SM} 

The SM can be tested at the quantum level with 
electroweak precision observables (EWPO), where many of them are related
to properties 
of the $Z$ and $W$ bosons. $Z$-boson properties are determined from measurements
of $e^+e^- \to f\bar{f}$ on the $Z$-pole.
A customary set of such (pseudo-)observables are the mass of the
$W$~boson, see the discussion in \refse{sec:intro}, as well as,
\begin{align}
A^f_{\rm FB} &= \frac{\sigma_f(\theta<\frac{\pi}{2})-
 \sigma_f(\theta>\frac{\pi}{2})}{\sigma_f(\theta<\frac{\pi}{2})+
 \sigma_f(\theta>\frac{\pi}{2})} \equiv \tfrac{3}{4}{\cal A}_e{\cal A}_f, 
 \label{eq:afb} \\[1ex]
A^f_{\rm LR} &= \frac{\sigma_f(P_e<0)-
 \sigma_f(P_e>0)}{\sigma_f(P_e<0)+
 \sigma_f(P_e>0)} \equiv {\cal A}_e|P_e|.
\end{align}
Here $\sigma_f$ denotes the cross section
$\sigma(e^+e^- \to f \bar f)$ measured at $\sqrt{s} = \MZ$,  
$\theta$ is the scattering angle and $P_e$ is the polarization of the
incoming electron beam.
The asymmetry parameters are commonly written as
\begin{align}
{\cal A}_f &= \frac{1-4|Q_f|\seff{f}}{1-4|Q_f|\seff{f}+8 (Q_f\seff{f})^2}.
\end{align}
Here $Q_f$ denotes the charge of the fermion, and $\seff{f}$ is the 
effective weak (fermionic) mixing angle.

\begin{figure*}[htb]
\centering
\includegraphics[width=135mm]{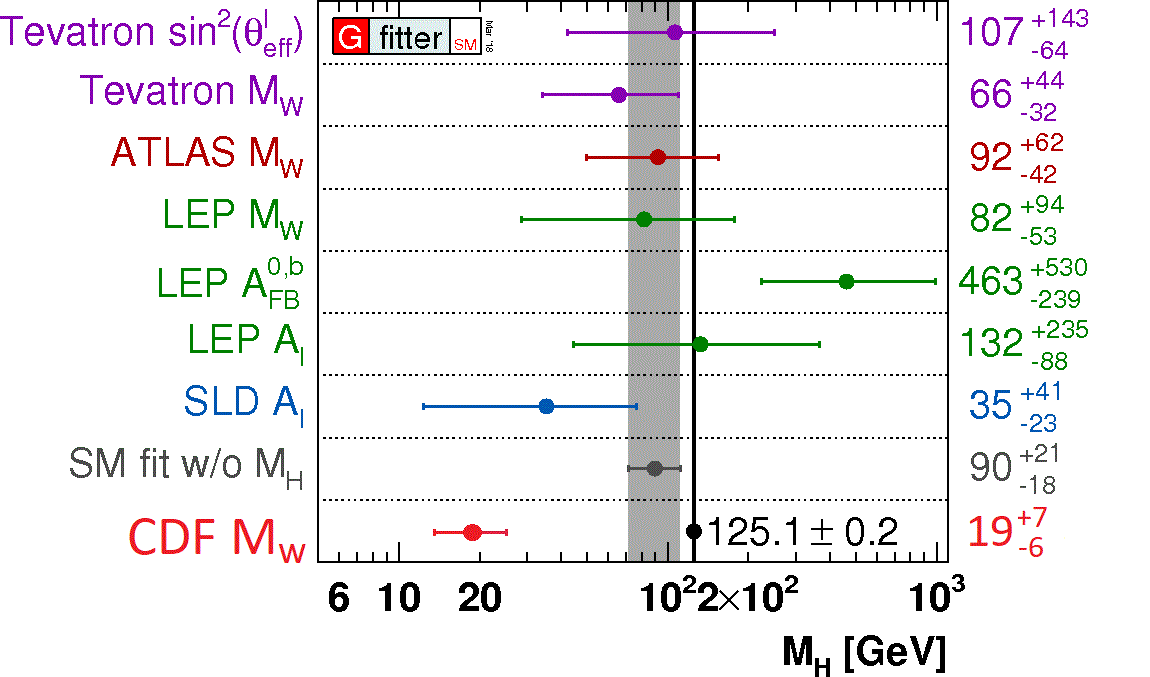}
\caption{Direct and indirect determinations of $\MH$ in the SM (see
  text). The figure (except the new result for \MWCDF) is taken from
  \citere{Haller:2018nnx}. }
\label{fig:MHSMwithCDF}
\end{figure*}

The theory prediction of an EWPO in the SM depends on the mass of the
SM Higgs boson, $\MH$, which enters via quantum
corrections~\cite{Sirlin:1980nh,Marciano:1980pb} (see
\citeres{Freitas:2019bre,Heinemeyer:2021rgq}
for related theory uncertainties).
The comparison of the experimental results for the EWPOs and their
predictions in the SM allows to extract a preferred range for $\MH$
from each observable. This is shown for seven different EWPOs in the
first seven lines of \reffi{fig:MHSMwithCDF}~\cite{Haller:2018nnx}.
Here it should be noted that the two most precise indirect
determinations, based on $A_{\rm FB}^b$ at LEP and
$A_l \equiv A_{\rm LR}^e$ at SLD disagree at the $3\,\si$ level.
The
average (without \MWCDF) as obtained by the GFitter collaboration is
given in the eighth line,
$\MH^{\rm fit} = 90^{+21}_{-18} \gev$~\cite{Haller:2018nnx}, which
has to be compared to the experimental value of~\cite{ATLAS:2015yey}
\begin{align}
  \MH^{\rm exp} = 125.1 \pm 0.2 \gev~,
\label{MHexp}
\end{align}
an agreement at the
$1.8\,\si$ level. Concerning the two most precise indirect
determinations of $\MH$ via $A_{\rm FB}^b$ and $A_{\rm LR}^e$, only
the average yields an acceptable agreement with $\MH^{\rm exp}$.
The last line of \reffi{fig:MHSMwithCDF} shows the
preferred $\MH$ range obtained from \MWCDF,
\begin{align}
  \MH^{\MWCDF{\rm -fit}} = 19^{+7}_{-6} \gev~.
  \label{MHMWCDF}
\end{align}
This value demonstrates the imcompatibility of the SM with the newly
measured value of \MWCDF.

\section{Extensions of the SM Higgs-boson sector}

While no conclusive signs of physics beyond the~SM (BSM) have been found so
far at the LHC, both the measurements of the properties of the Higgs
boson at $\sim 125\gev$ (its couplings are known up to now
to an experimental precision of roughly $10-20\%$) 
and the existing limits from the searches for new particles
leave significant room for BSM interpretations of the discovered Higgs-boson.
Extended Higgs-boson sectors naturally contain additional Higgs bosons with
masses larger than $125 \gev$. However, many extensions also offer the
possibilty of additional Higgs bosons that are {\em lighter} than
$125 \gev$.


\subsection{The 2HDM}

The Two Higgs Doublet Model (2HDM) extends the SM by a second Higgs
doublet (see \citere{Branco:2011iw} for a review).
The fields (after EWSB)
can be parametrized as
\begin{align}
\Phi_1 &= \left( \begin{array}{c} \phi_1^+ \\ \frac{1}{\sqrt{2}} (v_1 +
    \rho_1 + i \eta_1) \end{array} \right) \,, \nonumber \\
\Phi_2 &= \left( \begin{array}{c} \phi_2^+ \\ \frac{1}{\sqrt{2}} (v_2 +
    \rho_2 + i \eta_2) \end{array} \right) \;, 
\end{align}
where $\Phi_1$ and $\Phi_2$ are the two $SU(2)_L$ doublets with
hypercharge~1.
The parameters $v_1, v_2$ are the real vacuum expectation values
(vevs) acquired by the fields $\Phi_1$ and $\Phi_2$, respectively, 
with $\tb := v_2/v_1$ and $v^2 = v_1^2 + v_2^2$, where $v$ is the SM
vev.

In order to avoid the occurrence of tree-level flavor
changing neutral currents
(FCNC), a $Z_2$ symmetry is imposed on the
scalar potential, under which $\Phi_1 \to \Phi_1$ and $\Phi_2 \to -\Phi_2$.
This $Z_2$ symmetry, however, is softly broken by a bilinear term
usually written as $m_{12}^2 (\Phi_1^\dagger \Phi_2 + \mathrm{h.c.})$.
The extension of the $Z_2$ symmetry to the Yukawa
sector forbids tree-level FCNCs.
One can have four variants of the 2HDM, depending on the $Z_2$
parities of the fermions. The Higgs sector consists (after
diagonalization) of the light and heavy CP-even Higgses, $h$ and $H$,
the CP-odd Higgs, $A$, and a pair of charged Higgses, $H^\pm$.
The mixing angles $\al$ and $\be$ diagonalize the CP-even and -odd
part of the Higgs sector, respectively.
Here we assume that the light CP-even Higgs is identified with the
Higgs boson discovered at the LHC at $\sim 125 \gev$. 
Once $m_h$ and $v$ are set to $m_h \simeq 125 \gev$ and
$v\simeq 246 \gev$,  all four 2HDM types can be described in terms of
six input parameters, which are often chosen as: $\cos(\beta-\alpha)$,
$\tan\beta$,  $m_H$,  $m_A$, $m_{H^\pm}$ and $m_{12}$.  

After the CDF result for $\MW$ was published, many articles appeared
to describe the CDF value in BSM models, including analyses in the
2HDM, see \citeres{Song:2022xts,Bahl:2022xzi,Babu:2022pdn} for the
first papers. Here it should be noted that all these papers focused on
\MWCDF, \refeq{cdf-new}, but not on a possible new world average, see
\refeq{exp-nwe}. A first exception of an 2HDM analysis taking into
account a possible new world average can be found in
\citere{Arco:2022jrt}.

\begin{figure}[htb]
\centering
\includegraphics[width=80mm]{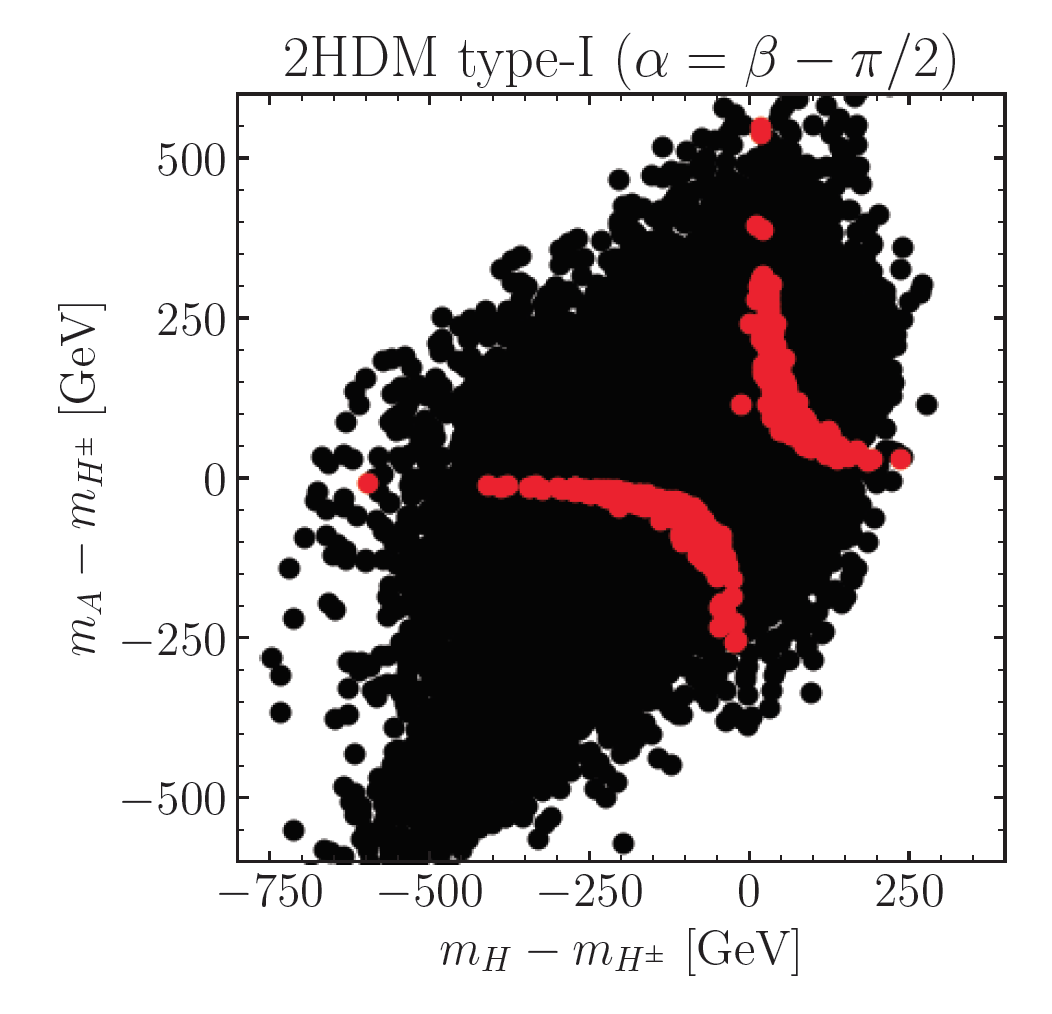}
\caption{
The $(m_H - m_{H^\pm})$-$(m_A - m_{H^\pm})$ plane in the 2HDM (type~I)
for $\cos(\be-\al) = 0$. Black points are the scanned parameter
points. Red points are within the $1\,\si$ interval of
\refeq{cdf-new}. Taken from \citere{Bahl:2022xzi}.
}
\label{fig:2HDM-DmHHp-DmAHp}
\end{figure}

Constraints from EWPOs on BSM models can in a simple approximation be
expressed in terms of the oblique parameters $S$, $T$ and
$U$~\cite{Peskin:1990zt,Peskin:1991sw}.
Effects from physics beyond the SM on these parameters can be significant
if the new physics contributions enter mainly through gauge boson
self-energies, as it is the case for extended Higgs sectors, and in
particular in the 2HDM.
Accordingly, the $W$-boson mass can be calculated
as a function of the oblique parameters, given
by~\cite{Grimus:2008nb}
\begin{align}
\MW^2 = \left. M_W^2 \right|_{\rm SM}
\left(
1 + \frac{\SW^2}{\CW^2 - \SW^2} \Delta r'
\right) \ ,
\label{MW-STU}
\end{align}
with
\begin{align}
\Delta r' = \frac{\alpha}{\SW^2}
\left(
-\frac{1}{2} S +
\CW^2 T +
\frac{\CW^2 - \SW^2}{4 \SW^2} U
\right) \ .
\label{Deltar-STU}
\end{align}
In the 2HDM the largest contribution enters via the $T$~parameter,
with a small correction from~$S$ and negligible contributions from~$U$.
The $T$~parameter is directly connected to the $\rho$-parameter via
$\al T \equiv \De\rho$, with
\begin{align}
  \De\rho &= \frac{\Sigma_Z^T(0)}{\MZ^2} -
  \frac{\Sigma_W^T(0)}{\MW^2}~,
\end{align}
where $\Sigma_{Z,W}^T(0)$ denotes the transversal part of the $Z$- or
$W$-boson self-energy at zero external momentum. 
The one-loop 2HDM BSM contributions to $\De\rho$ in
the alignment limit, $\cos(\be-\al) = 0$, are given
by~\cite{Hessenberger:2016atw}, 
\begin{align}
  \De\rho^{\rm 2HDM} &= \frac{\al}{16\,\pi^2\,\SW^2\,\MW^2}
  \Big\{ \frac{m_A^2 m_H^2}{m_A^2 - m_H^2} \log\frac{m_A^2}{m_H^2}
  \nonumber \\
  &-\frac{m_A^2 m_{H^\pm}^2}{m_A^2 - m_{H^\pm}^2} \log\frac{m_A^2}{m_{H^\pm}^2} 
\nonumber \\
  &-\frac{m_H^2 m_{H^\pm}^2}{m_H^2 - m_{H^\pm}^2} \log\frac{m_H^2}{m_{H^\pm}^2} 
  + m_{H^\pm}^2 \Big\}~. 
\end{align}
It vanishes for $m_A = m_{H^\pm}$ or $m_H = m_{H^\pm}$. 
Large positive contributions, required to accomodate \MWCDF, are found
for large mass splittings and
\begin{align}
  (m_H - m_{H^\pm})(m_A - m_{H^\pm}) > 0~.
\end{align}
This is demonstrated in
\reffi{fig:2HDM-DmHHp-DmAHp}~\cite{Bahl:2022xzi}, where the plane
$(m_H - m_{H^\pm})$-$(m_A - m_{H^\pm})$ is shown in the 2HDM (type~I)
for $\cos(\be-\al) = 0$. The black points are the scanned parameter
points. Shown in red are the points within the $1\,\si$ interval of
\refeq{cdf-new}. 

Using the approximation of the $T$~parameter, the corrections to the
effective weak leptonic mixing angle can be expressed as
\begin{align}
\sweff = \left. \sweff \right|_{\rm SM}
- \alpha \frac{\CW^2 \SW^2}{\CW^2 - \SW^2} T \ .
\end{align}
Because of the different sign of the $T$~parameter contribution
w.r.t.\ \refeq{Deltar-STU} a positive contribution to the prediction
of $\MW$ corresponds to a negative contribution to $\sweff$.


\subsection{The N2HDM}

There are several excesses in the searches for light Higgs-boson
around $\sim 95 \gev$. 
Results based on the first year of CMS~Run\,2 data for Higgs-boson
searches in the diphoton
final state show a local excess of about~$3\,\sigma$ at a mass of
$95\gev$~\cite{CMS:2018cyk}, where
a similar excess of~$2\,\sigma$ occurred in the Run\,1 data at a comparable
mass~\cite{CMS:2015ocq}.
Combining 7, 8~and first year $13 \tev$ data
(and assuming that the $gg$ production dominates)
the excess is most pronounced at a mass of $95.3 \gev$ with a local
significance of $2.8\,\sigma$.
First Run\,2~limits from~ATLAS
with~$80$\,fb$^{-1}$ in the~$\gamgam$~searches below~$125$\,GeV were
reported in 2018~\cite{ATLAS:2018xad} and are
substantially weaker than the corresponding upper limit obtained by CMS
at and around $95 \gev$.

CMS recently published the results for
the search for additional Higgs bosons in the $\tautau$
channel~\cite{CMS:2022rbd}. 
Utilizing the full Run~2 data set, 
in \citere{CMS:2022rbd} the CMS collaboration
reported an excess in the low-mass region for the gluon-fusion production
mode and subsequent decay into $\tautau$ pairs that is compatible with the 
excess that has been observed by CMS in the diphoton search.
For a mass value of $95\gev$ CMS reports a local significance of $2.6\,\sigma$.
Up to now there exists no corresponding 
result for the low-mass search in the $\tautau$ final state from the ATLAS 
collaboration in this mass range.

Searches for a low-mass Higgs boson that were previously carried out at 
LEP resulted in a $2.3\,\sigma$ local excess
observed in the~$e^+e^-\to Z(H\to b\bar{b})$
searches~\cite{Barate:2003sz}
at a mass of
about $98 \gev$; due to the $b \bar b$ final state the
mass resolution was rather coarse.
Because of this limited mass resolution in the
$\bbbar$ final state at LEP  
this excess can be compatible with the slightly lower mass of $95 \gev$,
where the two CMS excesses have been observed.

In \citere{Biekotter:2022jyr} it was shown that the three excesses can
be described consistently in the N2HDM (the Two-Higgs-Doublet Model with an
additional real singlet~\cite{Chen:2013jvg,Muhlleitner:2016mzt}). In
the N2HDM, contrary to the 2HDM, three CP-even Higgs bosons are
present. The three exceeses can be accomodates simultaneously in the
N2HDM of Yukawa type~IV by the lightest CP-even Higgs boson, $h_{95}$,
with a mass of $\sim 95 \gev$, which has a large singlet component. At
the same time the second lightest CP-even Higgs boson with a mass of
$\sim 125 \gev$, $h_{125}$, agrees well with the LHC measurements, and
the rest of the Higgs sector is in agreement with the Higgs-boson
searches at LEP and the LHC.

\begin{figure}[htb]
\centering
\includegraphics[width=80mm]{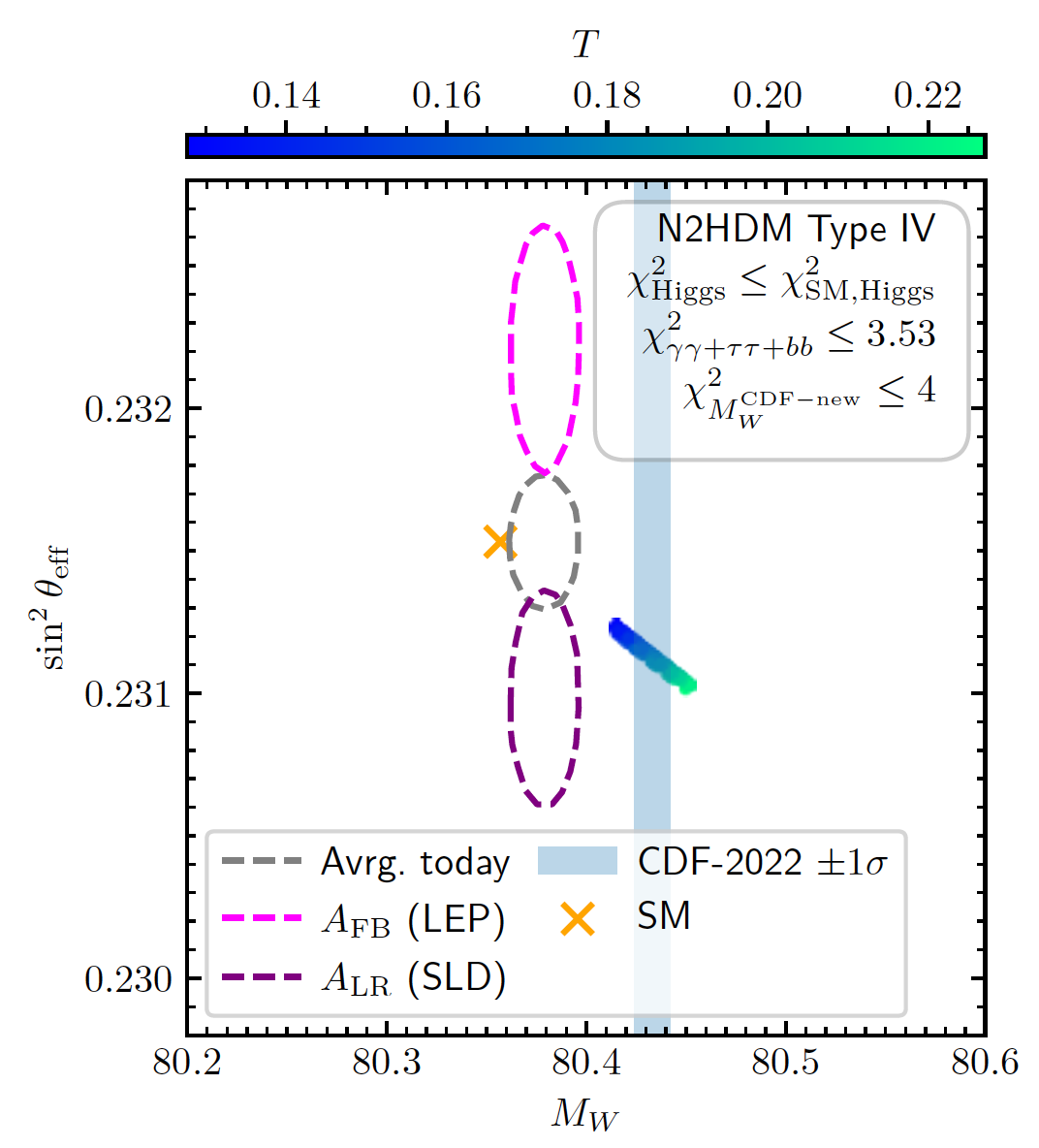}
\caption{
  The $\MW$-$\sweff$ plane in the 2HDM in type~IV (see text).
 Taken from \citere{Biekotter:2022abc}.
}
\label{fig:MW-SW}
\end{figure}

In \citere{Biekotter:2022abc} it was subsequently demonstrated that
this model can also ``comfortably'' accomodate \MWCDF.
In \reffi{fig:MW-SW}~\cite{Biekotter:2022abc} we show the predictions
for $\MW$ and $\sweff$ in the N2HDM type~IV points that fit well the
$\ga\ga$, $\tau\tau$ and $b\bar b$ excesses via $h_{95}$, while the
$h_{125}$ is in good agreement with the LHC Higgs-boson rate
measurements, see \citere{Biekotter:2022abc} for details.
The color coding of the points indicates the
value of $T$. The light blue regions corresponds
to the new CDF measurement within $\pm 1 \ \sigma$.
The purple and the magenta ellipses indicate the $68\%$ confidence
level limits from the two individually most precise 
measurements of $\sweff$ via $A_{\rm FB}$ at LEP and
$A_{\rm LR}$ at SLD, respectively, whereas the gray ellipse
indicates the PDG average.
The orange cross indicates the SM prediction.
One can see that the parameter points that fit the
new CDF measurement of the $W$-boson mass feature
also sizable modifications of $\sweff$
compared to the SM prediction, as mentioned above.
The values of $\sweff$ featured in the parameter points of the scan are smaller
than the SM value, not touching the current $1\,\sigma$
ellipse. However, as discussed above, the PDG average is composed of
two measurements that are compatible only at the $\sim 3\,\sigma$
level: the one using the forward-backward asymmetry in $e^+e^- \to b
\bar b$ measured at LEP~\cite{ALEPH:2005ab}, and the one obtained from
the left-right asymmetry in $e^+e^- \to e^+e^-$ measured at
SLD~\cite{ALEPH:2005ab}. It can be observed that the data points
preferred by the $\MW$ measurement of CDF  are in better 
agreement with the SLD measurement based on
$A_{\rm LR}^e$, whereas the tension increases with
the value of $\sin^2\theta_{\rm eff}$ extracted
at LEP based on measurements of $A_{\rm FB}^b$.

The correlation between the effective weak
mixing angle and the mass of the $W$~boson
is expected to arise generically in models in
which the new CDF measurement of $M_W$ is
accommodated mainly via the breaking of the
custodial symmetry by means of a non-zero
$T$~parameter (and not via, e.g., BSM vertex and box
contributions to the muon decay).
The presence of the additional singlet state of the N2HDM compared to the
2HDM has no sizable impact on the distribution
of parameter points in \reffi{fig:MW-SW} in the investigated
scenario, because the mixing between the
singlet-like state $h_{95}$ and the heavy CP-even
Higgs boson is very small as a result
of the large mass difference.


\section{The MSSM}

The $W$-boson mass can be calculated in the Minimal Supersymmetric
Standard Model (MSSM) by evaluating the supersymmetric (SUSY)
contributions to $\De r$, see \refeq{eq:mwpred}. The most precise
prediction for $\MW$ in the MSSM is based on the full one-loop result for
$\De r$~\cite{Heinemeyer:2006px,Heinemeyer:2007bw,Heinemeyer:2013dia}
(see also \citere{Chankowski:1993eu}),
supplemented by the leading two-loop
corrections~\cite{Djouadi:1996pa,Djouadi:1998sq,Haestier:2005ja}.
The leading one- and two-loop contributions arise from isospin splitting
between different SUSY particles and enter via the quantity $\De\rho$.
At the one-loop level the squarks enter only via self-energy contributions,
i.e.\ predominantly via $\De\rho$. The same is true for the corresponding
contribution of pure slepton loops, while the contributions of the chargino and
neutralino sector enter also via vertex and box diagrams. In our MSSM prediction
for $\MW$ the contributions involving SUSY particles are combined with all
available SM-type contributions up to the four-loop level as described above.
Technically this is done by using a fit formula for the SM
contributions beyond one-loop for $\De r$~\cite{Awramik:2003rn}.
This ensures that the
state-of-the-art SM prediction is recovered in the decoupling limit where all
SUSY mass scales are heavy.

In
\citeres{Chakraborti:2020vjp,Chakraborti:2021kkr,Chakraborti:2021dli,Chakraborti:2021mbr}
the EW sector of the MSSM was analyzed
taking into account all relevant experimental
data, i.e.\ data that is directly connected to the EW sector.
It was assumed that the Lightest
SUSY Particle (LSP)  is the lightest neutralino,~$\neu{1}$, 
as the Dark Matter (DM) candidate of the model.
The experimental results employed in the analyses 
comprise the direct searches at the
LHC~\cite{ATLAS-SUSY,CMS-SUSY}, the DM relic abundance~\cite{Planck:2018vyg},
the DM direct detection
experiments~\cite{XENON:2018voc,LUX:2016ggv,PandaX-II:2017hlx}
together with the deviation on the value
of the anomalous magnetic moment of the muon, $\amu$. The comparison
of the SM prediction~\cite{Aoyama:2020ynm} with the experimental
result~\cite{Muong-2:2006rrc,Muong-2:2021ojo} yields
\begin{align}
\Delta\amu \equiv \amu^{\rm exp} - \amu^{\rm SM}
&= (25.1 \pm 5.9) \times 10^{-10}~,
\label{gmt-diff}
\end{align}
corresponding to a $4.2\,\si$ discrepancy.

In
\citeres{Chakraborti:2020vjp,Chakraborti:2021kkr,Chakraborti:2021dli,Chakraborti:2021mbr}
five different scenarios were identified that are in agreement with
the above listed limits, classified by the mechanism that has the main
impact on the resulting LSP relic density.
The scenarios differ by the nature of the Next-to-LSP
(NLSP). They comprise $\cha1$-coannhiliation, $\Slpm$-coannihilation
with either 
``left-'' or ``right-handed'' sleptons close in mass to the
LSP (``case-L'' and ``case-R'', respectively), wino DM, as well as
higgsino DM.
In the first
three scenarios the full amount of DM can be provided by the MSSM,
whereas in the latter two cases the measured DM density serves as an
upper limit. Requiring \refeq{gmt-diff} at the $2\,\si$ level,
together with the collider and DM constraints,
results in upper limits on the LSP masses at the level of $\sim 500 \gev$
to $\sim 600 \gev$ for all five scenarios.
Corresponding upper limits on the mass of the NLSP are obtained for
only slightly higher mass values.

\begin{figure*}[htb]
\centering
\includegraphics[width=135mm]{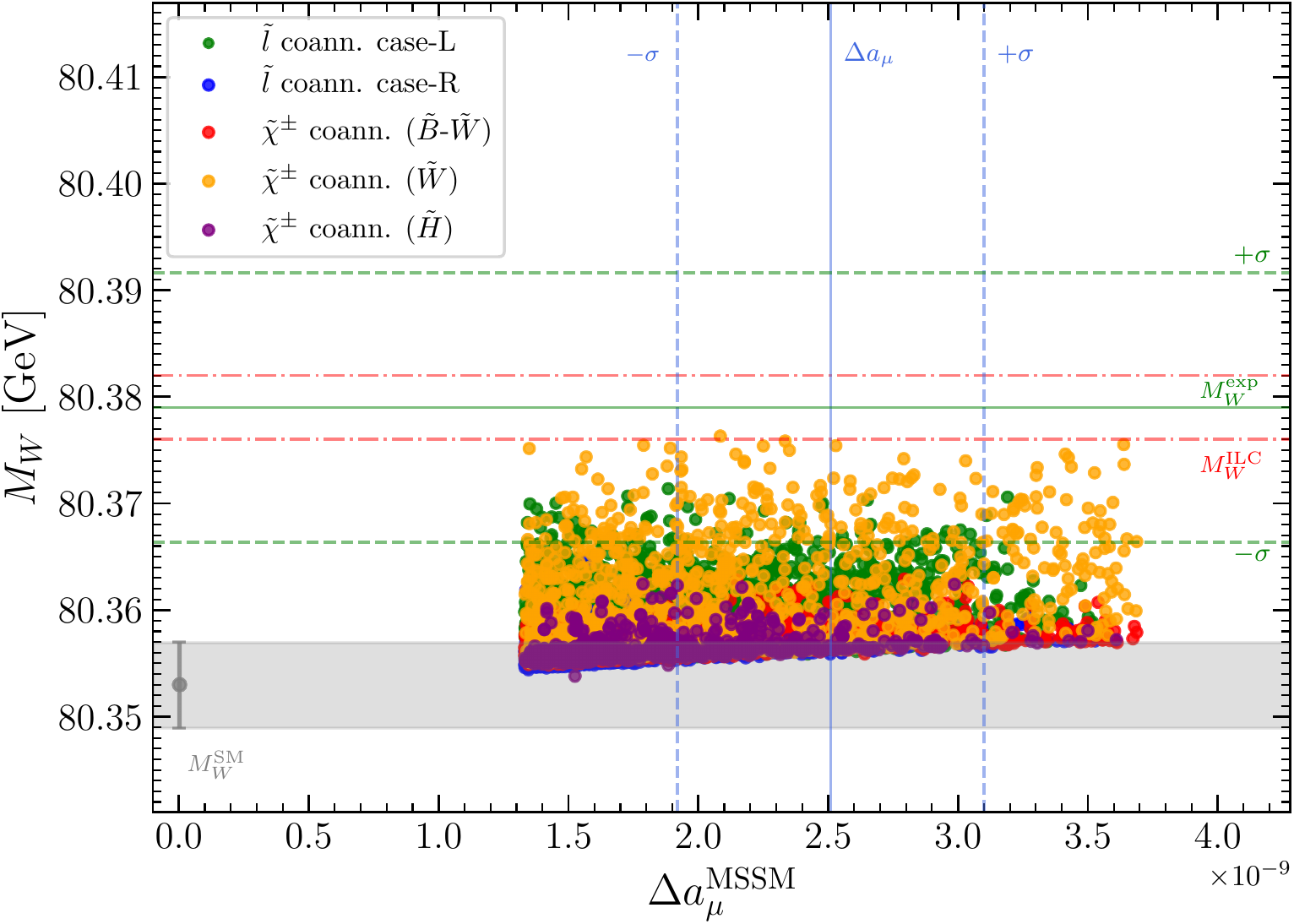}
\caption{The $\De\amu^{\rm MSSM}$-$\MW$ plane with the results for the
  five considered scenarios (see text).
  Taken from \citere{Bagnaschi:2022qhb}. }
\label{fig:amu-MW}
\end{figure*}

In \citere{Bagnaschi:2022qhb} the 
contributions to the $\MW$ prediction from the EW MSSM in the five
scenarios that are in agreement with all experimental data and in
particular with \refeq{gmt-diff} were analyzed. The results are shown
in \reffi{fig:amu-MW} in the $\amu^{\rm MSSM}$-$\MW$ plane. 
The prediction for $\De\amumssm$ has been evaluated with the code
{\tt GM2Calc-1.7.5}~\cite{Athron:2015rva}.
The vertical solid blue line indicates the value of
$\De\amu$ as given in \refeq{gmt-diff}, while its
$\pm 1\,\si$ range is indicated by the blue dashed vertical lines.
The displayed points are restricted to the $\pm 2\,\si$ range of $\De\amu$.
The horizontal lines indicate the old central value for $\MW^{\rm exp}$
(solid green), together with its $\pm 1\,\si$ uncertainties (green dashed)
and the anticipated ILC $\pm 1\,\si$ (red dot-dashed) uncertainties.
The SM prediction is shown in gray, including the theoretical uncertainty
from unknown higher-order corrections.

One can observe that relatively light SUSY particles that are required
for larger values 
of $\De\amumssm$ give rise to a slight increase in the prediction for
$\MW$ that is independent of the variation of the other parameters in
the scan. While this lower limit on the predicted value of $\MW$ is very
similar in the five DM scenarios, there are important differences in
the highest $\MWMSSM$ values that are reached.
The largest predicted values of $\MWMSSM$, nearly reaching the ``old''
central value of $\MWexp$, are obtained for the wino DM case.
Accordingly, for the wino DM case the electroweak sector of the MSSM behaves
in such a way that the predicted values for $\MW$ and the anomalous magnetic
moment of the muon can simultaneously be very close to the experimental
central values, while respecting all other constraints on the model.
For the $\Slpm$-coannihilation case-L the highest obtained $\MWMSSM$ values
are somewhat lower but still within the $\pm 1\,\si$ range of
$\MWexp$. This analysis demonstrates that the EW sector can yield an
increase of $\sim 20 \mev$ to the $\MW$ prediction, but cannot yield
values close to the new possible world average, \refeq{exp-nwe}.

\begin{figure}[htb]
\centering
\includegraphics[width=80mm]{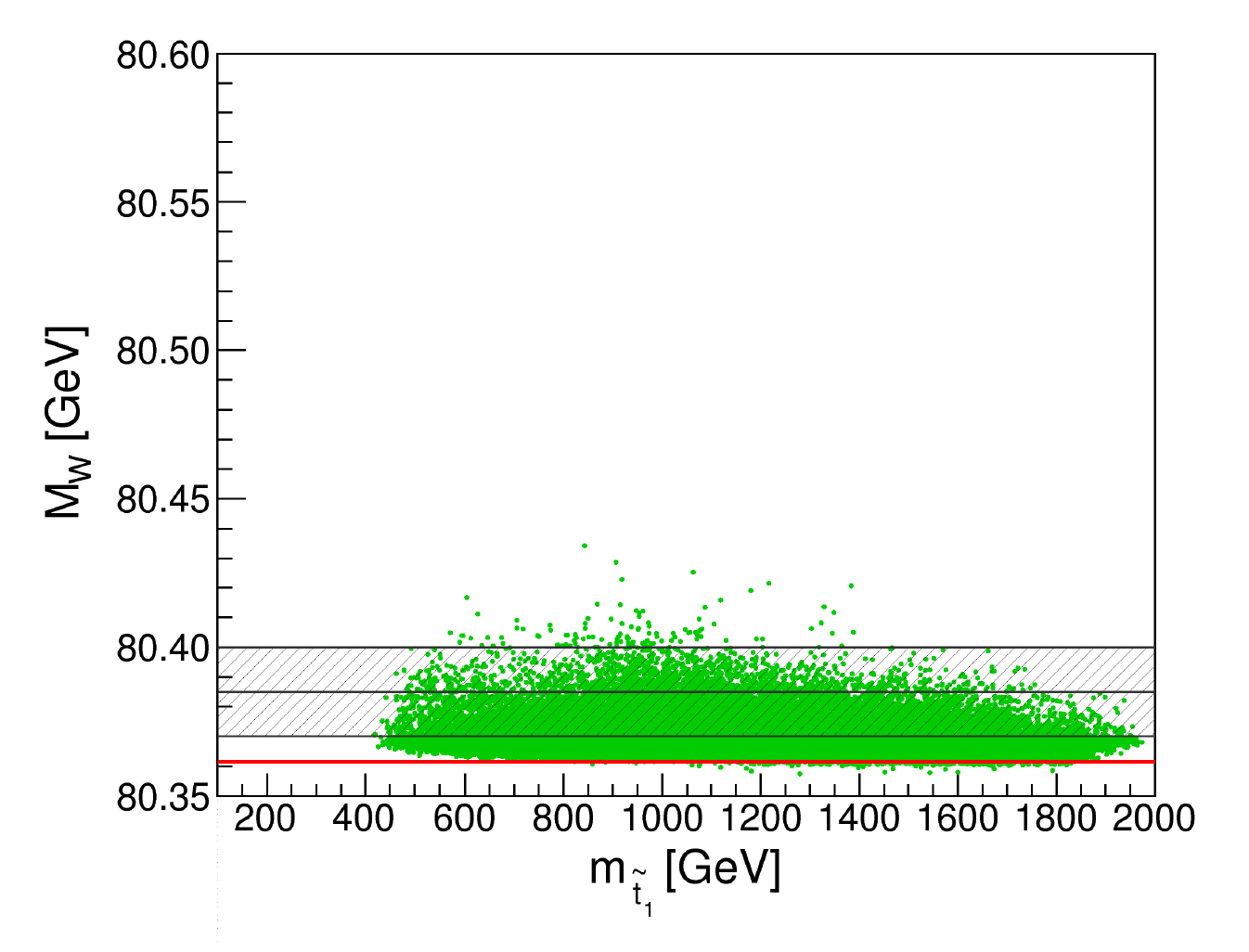}
\caption{
  The $m_{\tilde t_1}$-$\MW$ plane in the MSSM (see text).
 Taken from \citere{Heinemeyer:2013dia}.
}
\label{fig:mstop1-MW}
\end{figure}

The result shown in \reffi{fig:amu-MW}, however, does not imply that a
possible new world average of $\MWexp$ as given in \refeq{exp-nwe}
cannot be reached, as it focused on the contributions from the EW
sector of the MSSM. Large corrections to $\De\rho$ can originate in the
stop/sbottom sector (the SUSY partners of the tops and bottoms), which
has been analyzed in \citere{Heinemeyer:2013dia}. 
The results of an extensive MSSM parameter scan are shown as green
points in the $m_{\tilde t_1}$-$\MW$ plane in
\reffi{fig:mstop1-MW}~\cite{Heinemeyer:2013dia}. The gray horizontal
band indicates the then current experimental $\pm 1\,\si$ range of
$\MWexp$, and the red line corresponds to the SM prediction. Squarks
and gluinos have been chosen above the TeV scale to avoid LHC limits, and the
lightest sbottom mass is heavier than $500 \gev$. Furthermore, a limit
of $2/5 \le m_{\tilde t_i}/m_{\tilde b_j} \le 5/2$ ($i,j = 1,2$) has
been applied. One can observe that for light stops above the TeV scale
(very roughly corresponding to current LHC limits) indeed points in the range
of a possible new world averange, \refeq{exp-nwe}, are found. It
should furthermore be noted that even larger values could be obtained
by dropping the condition on the ratio of stop and sbottom masses. 

\section{Conclusions}

The CDF collaboration recently reported a measurement of the $W$-bosos
mass, $M_W$, showing a large positive
deviation from the SM prediction.
The question arises whether extensions of the SM exist that can
accommodate such large values, and what further phenomenological
consequences arise from this.
Here we first briefly summarized of the implications of the new CDF measurement
on the SM. We demonstrated that the theory prediction of $\MW$ in the
SM yields an indirect determination of
$\MH^{\MWCDF{\rm -fit}} = 19^{+7}_{-6} \gev$, far below the experimental value.

We briefly reviewed the predictions of $\MW$ in the 2HDM and showed
that it is easy to reach a \MWCDF\ by an increased mass splitting
between the heavy Higgs bosons (assuming that the light CP-even Higgs
boson corresponds to the state discovered at $\sim 125 \gev$). The
contribution of the additional Higgses, entering dominantly via
$\De\rho$, on the other hand lower the prediction of the effective
weak leptonic mixing angle, $\sweff$. For a 2HDM parameter point yielding
$\sim \MWCDF$ the $\sweff$ prediction would be incompatible with the
determination via $A_{\rm FB}^b$ at LEP, and also with the current
world average, but would be in agreement with the $\sweff$
determination via $A_{\rm LR}^e$ at SLD.
Going to the N2HDM, it was shown that parameter points that give a
good fit to three independent excesses in the low-mass Higgs-boson
searches at $\sim 95 \gev$ can also accomodate ``comfortably'' \MWCDF.

Finally we reviewd the $\MW$ prediction in the MSSM. The EW sector of
the MSSM, taking into account all relevant constraints (\gmin2, DM
relic density, DM direct detection limits as well as LHC searches) can
shift the $\MW$ prediction upwards by ut to $\sim 20 \mev$ w.r.t.\ the
SM prediction. Taking into account also contributions from the
stop/sbottom sector values around a possible new world average of
$\MWexp$, \refeq{exp-nwe}, can be reached.


\subsection*{Acknowledgements}

I thank my collaborators with whom some of the results reviewed here
have been obtained:
E.~Bagnaschi,
T.~Biek\"otter,
M.~Chakraborti,
W.~Hollik,
I.~Saha,
C.~Schappacher, 
G.~Weiglein
and
L.~Zeune.
I thank
M.~Berger
and
M.~Martinez
for help with \reffi{fig:MHSMwithCDF}.
The work of S.H.\ is supported in part by
the grant PID2019-110058GB-C21 funded by
``ERDF A way of making Europe'' and by
MCIN/AEI/10.13039/501100011033, and in part
by the grant CEX2020-001007-S funded by
MCIN/AEI/10.13039/501100011033.



\begin{thebibliography}{99}   


\bibitem{Sirlin:1980nh}
  A.~Sirlin,
  Phys. Rev. D \textbf{22} (1980), 971-981.

\bibitem{Marciano:1980pb}
  W.~J.~Marciano and A.~Sirlin,
  Phys. Rev. D \textbf{22} (1980), 2695
  [erratum: Phys. Rev. D \textbf{31} (1985), 213].

\bibitem{PDG2022}
  R.~L.~Workman \textit{et al.} [Particle Data Group],
  PTEP \textbf{2022} (2022), 083C01.

\bibitem{CDF:2022hxs}
  T.~Aaltonen \textit{et al.} [CDF],
  Science \textbf{376} (2022) no.6589, 170-176.

\bibitem{Freitas:2019bre}
  A.~Freitas, S.~Heinemeyer, M.~Beneke, A.~Blondel, S.~Dittmaier, J.~Gluza,
  A.~Hoang, S.~Jadach, P.~Janot and J.~Reuter, \textit{et al.}
  [arXiv:1906.05379 [hep-ph]].

\bibitem{Heinemeyer:2021rgq}
  S.~Heinemeyer, S.~Jadach and J.~Reuter,
  Eur. Phys. J. Plus \textbf{136} (2021) no.9, 911
  [arXiv:2106.11802 [hep-ph]].
  
\bibitem{Haller:2018nnx}
  J.~Haller, A.~Hoecker, R.~Kogler, K.~M\"onig, T.~Peiffer and J.~Stelzer,
  Eur. Phys. J. C \textbf{78} (2018) no.8, 675
  [arXiv:1803.01853 [hep-ph]].

\bibitem{ATLAS:2015yey}
  G.~Aad \textit{et al.} [ATLAS and CMS],
  Phys. Rev. Lett. \textbf{114} (2015), 191803
  [arXiv:1503.07589 [hep-ex]].

\bibitem{Branco:2011iw}
G.~C.~Branco, P.~M.~Ferreira, L.~Lavoura, M.~N.~Rebelo, M.~Sher and
J.~P.~Silva,
Phys. Rept. \textbf{516} (2012), 1-102
[arXiv:1106.0034 [hep-ph]].

\bibitem{Song:2022xts}
  H.~Song, W.~Su and M.~Zhang,
  surements,''
  [arXiv:2204.05085 [hep-ph]].

\bibitem{Bahl:2022xzi}
  H.~Bahl, J.~Braathen and G.~Weiglein,
  [arXiv:2204.05269 [hep-ph]].

\bibitem{Babu:2022pdn}
  K.~S.~Babu, S.~Jana and Vishnu~P.~K.,
  [arXiv:2204.05303 [hep-ph]].
  
\bibitem{Arco:2022jrt}
F.~Arco, S.~Heinemeyer and M.~J.~Herrero,
[arXiv:2207.13501 [hep-ph]].

\bibitem{Peskin:1990zt}
M.~E.~Peskin and T.~Takeuchi,
Phys. Rev. Lett. \textbf{65} (1990), 964-967.

\bibitem{Peskin:1991sw}
M.~E.~Peskin and T.~Takeuchi,
Phys. Rev. D \textbf{46} (1992), 381-409.

\bibitem{Grimus:2008nb}
W.~Grimus, L.~Lavoura, O.~M.~Ogreid and P.~Osland,
Nucl. Phys. B \textbf{801} (2008), 81-96
[arXiv:0802.4353 [hep-ph]].

\bibitem{Hessenberger:2016atw}
S.~Hessenberger and W.~Hollik,
Eur. Phys. J. C \textbf{77} (2017) no.3, 178
[arXiv:1607.04610 [hep-ph]].

\bibitem{CMS:2018cyk}
A.~M.~Sirunyan \textit{et al.} [CMS],
Phys. Lett. B \textbf{793} (2019), 320-347
[arXiv:1811.08459 [hep-ex]].

\bibitem{CMS:2015ocq}
 [CMS],
 CMS-PAS-HIG-14-037.

\bibitem{ATLAS:2018xad}
 [ATLAS],
 ATLAS-CONF-2018-025.

\bibitem{CMS:2022rbd}
 [CMS],
 CMS-PAS-HIG-21-001.

\bibitem{Barate:2003sz}
R.~Barate \textit{et al.} [LEP Working Group for Higgs boson searches, ALEPH,
DELPHI, L3 and OPAL],
Phys. Lett. B \textbf{565} (2003), 61-75
[arXiv:hep-ex/0306033 [hep-ex]].

\bibitem{Biekotter:2022jyr}
T.~Biek\"otter, S.~Heinemeyer and G.~Weiglein,
[arXiv:2203.13180 [hep-ph]].

\bibitem{Chen:2013jvg}
C.~Y.~Chen, M.~Freid and M.~Sher,
Phys. Rev. D \textbf{89} (2014) no.7, 075009
[arXiv:1312.3949 [hep-ph]].

\bibitem{Muhlleitner:2016mzt}
M.~Muhlleitner, M.~O.~P.~Sampaio, R.~Santos and J.~Wittbrodt,
JHEP \textbf{03} (2017), 094
[arXiv:1612.01309 [hep-ph]].

\bibitem{Biekotter:2022abc}
T.~Biek\"otter, S.~Heinemeyer and G.~Weiglein,
[arXiv:2204.05975 [hep-ph]].

\bibitem{ALEPH:2005ab}
S.~Schael \textit{et al.} [ALEPH, DELPHI, L3, OPAL, SLD, LEP Electroweak
Working Group, SLD Electroweak Group and SLD Heavy Flavour Group],
Phys. Rept. \textbf{427} (2006), 257-454
[arXiv:hep-ex/0509008 [hep-ex]].

\bibitem{Heinemeyer:2006px}
S.~Heinemeyer, W.~Hollik, D.~Stockinger, A.~M.~Weber and G.~Weiglein,
JHEP \textbf{08} (2006), 052
[arXiv:hep-ph/0604147 [hep-ph]].

\bibitem{Heinemeyer:2007bw}
S.~Heinemeyer, W.~Hollik, A.~M.~Weber and G.~Weiglein,
JHEP \textbf{04} (2008), 039
[arXiv:0710.2972 [hep-ph]].

\bibitem{Heinemeyer:2013dia}
S.~Heinemeyer, W.~Hollik, G.~Weiglein and L.~Zeune,
JHEP \textbf{12} (2013), 084
[arXiv:1311.1663 [hep-ph]].

\bibitem{Chankowski:1993eu}
P.~H.~Chankowski, A.~Dabelstein, W.~Hollik, W.~M.~Mosle, S.~Pokorski and
J.~Rosiek,
Nucl. Phys. B \textbf{417} (1994), 101-129.

\bibitem{Djouadi:1996pa}
A.~Djouadi, P.~Gambino, S.~Heinemeyer, W.~Hollik, C.~Junger and G.~Weiglein,
Phys. Rev. Lett. \textbf{78} (1997), 3626-3629
[arXiv:hep-ph/9612363 [hep-ph]].

\bibitem{Djouadi:1998sq}
A.~Djouadi, P.~Gambino, S.~Heinemeyer, W.~Hollik, C.~Junger and G.~Weiglein,
Phys. Rev. D \textbf{57} (1998), 4179-4196
[arXiv:hep-ph/9710438 [hep-ph]].

\bibitem{Haestier:2005ja}
J.~Haestier, S.~Heinemeyer, D.~Stockinger and G.~Weiglein,
JHEP \textbf{12} (2005), 027
[arXiv:hep-ph/0508139 [hep-ph]].

\bibitem{Awramik:2003rn}
M.~Awramik, M.~Czakon, A.~Freitas and G.~Weiglein,
Phys. Rev. D \textbf{69} (2004), 053006
[arXiv:hep-ph/0311148 [hep-ph]].

\bibitem{Chakraborti:2020vjp}
M.~Chakraborti, S.~Heinemeyer and I.~Saha,
Eur. Phys. J. C \textbf{80} (2020) no.10, 984
[arXiv:2006.15157 [hep-ph]].

\bibitem{Chakraborti:2021kkr}
M.~Chakraborti, S.~Heinemeyer and I.~Saha,
Eur. Phys. J. C \textbf{81} (2021) no.12, 1069
[arXiv:2103.13403 [hep-ph]].

\bibitem{Chakraborti:2021dli}
M.~Chakraborti, S.~Heinemeyer and I.~Saha,
Eur. Phys. J. C \textbf{81} (2021) no.12, 1114
[arXiv:2104.03287 [hep-ph]].

\bibitem{Chakraborti:2021mbr}
M.~Chakraborti, S.~Heinemeyer, I.~Saha and C.~Schappacher,
Eur. Phys. J. C \textbf{82} (2022) no.5, 483
[arXiv:2112.01389 [hep-ph]].

\bibitem{ATLAS-SUSY}
See: {\tt https://twiki.cern.ch/twiki/bin/view/\\
AtlasPublic/SupersymmetryPublicResults}~.

\bibitem{CMS-SUSY}
See: {\tt https://twiki.cern.ch/twiki/bin/view/\\
CMSPublic/PhysicsResultsSUS}~.

\bibitem{Planck:2018vyg}
N.~Aghanim \textit{et al.} [Planck],
Astron. Astrophys. \textbf{641} (2020), A6
[erratum: Astron. Astrophys. \textbf{652} (2021), C4]
[arXiv:1807.06209 [astro-ph.CO]].

\bibitem{XENON:2018voc}
E.~Aprile \textit{et al.} [XENON],
Phys. Rev. Lett. \textbf{121} (2018) no.11, 111302
[arXiv:1805.12562 [astro-ph.CO]].

\bibitem{LUX:2016ggv}
D.~S.~Akerib \textit{et al.} [LUX],
Phys. Rev. Lett. \textbf{118} (2017) no.2, 021303
[arXiv:1608.07648 [astro-ph.CO]].

\bibitem{PandaX-II:2017hlx}
X.~Cui \textit{et al.} [PandaX-II],
Phys. Rev. Lett. \textbf{119} (2017) no.18, 181302
[arXiv:1708.06917 [astro-ph.CO]].

\bibitem{Aoyama:2020ynm}
T.~Aoyama, N.~Asmussen, M.~Benayoun, J.~Bijnens, T.~Blum, M.~Bruno,
I.~Caprini, C.~M.~Carloni Calame, M.~C\`e and G.~Colangelo, \textit{et al.}
Phys. Rept. \textbf{887} (2020), 1-166
[arXiv:2006.04822 [hep-ph]].

\bibitem{Muong-2:2006rrc}
G.~W.~Bennett \textit{et al.} [Muon g-2],
Phys. Rev. D \textbf{73} (2006), 072003
[arXiv:hep-ex/0602035 [hep-ex]].

\bibitem{Muong-2:2021ojo}
B.~Abi \textit{et al.} [Muon g-2],
Phys. Rev. Lett. \textbf{126} (2021) no.14, 141801
doi:10.1103/PhysRevLett.126.141801
[arXiv:2104.03281 [hep-ex]].


\bibitem{Bagnaschi:2022qhb}
E.~Bagnaschi, M.~Chakraborti, S.~Heinemeyer, I.~Saha and G.~Weiglein,
Eur. Phys. J. C \textbf{82} (2022) no.5, 474
[arXiv:2203.15710 [hep-ph]].

\bibitem{Athron:2015rva}
P.~Athron, M.~Bach, H.~G.~Fargnoli, C.~Gnendiger, R.~Greifenhagen, J.~h.~Park,
S.~Pa\ss{}ehr, D.~St\"ockinger, H.~St\"ockinger-Kim and A.~Voigt,
Eur. Phys. J. C \textbf{76} (2016) no.2, 62
[arXiv:1510.08071 [hep-ph]].
  

\end{thebibliography}
\end{document}